# Knowledge and skills requirements for the software design and testing of automotive applications

J. Zabavnik, A. Riel, M. Marguč, M. Rodič

*Abstract*— The required knowledge and skills that should be provided to the novice developer, designing and testing the safety critical device in automotive industry using Hardware-in-the-Loop (HiL), are presented in the paper. They should be available to the student finishing the MSc level of Electrical Engineering or Mechatronics that aims to seek the employment in the automotive industry. The development process is presented in short, together with the brief explanations of phases, which include some typical examples of written text in the documentation (requirements, test cases, etc.). The process follows the Automotive SPICE and focuses on the tip of the V-model. The presented text can serve as a useful information in the process of adapting the existing curriculum to the new occurring needs introduced by the new development and testing processes in industry.

*Index Terms*— software engineering, requirements, Hardware-in-the-Loop, testing, automotive

## I. Introduction

IN the modern vehicles there is a growing number of electric, electronic and programmed electronic devices. This is not only the case for the main drive (drivetrain), but also for a big number of auxiliary devices. In many cases previously mechanic or hydraulic devices are being replaced with the electric ones. For many such devices (so-called safety critical devices), failures could result in serious, even fatal, injuries of the passenger in the vehicle. These failures must be avoided as far as possible at reasonable cost and in a way sustainable for the producer of the vehicle.

Using extremely reliable devices would in many cases result in a very high cost of the final product, making it unaffordable for the customer. The solution is therefore to develop in such a way, that the devices remain safe even in the case of their failure, i.e., to achieve functional safety of the device and as a result also vehicle. Failures shall be avoided, however if this is not possible, their effects have to be mitigated. This goal can be achieved with the use of fault-tolerant devices developed according to the proper development cycle.

The development cycle in the automotive industry is based on the classical V-model, where the development from the user requirements to the validation of the product is described. In automotive industry processes are specified in compliance with the Automotive SPICE standard [1]. Both the system and software development process is upgraded with the processes and methods prescribed in IEC/ISO 26262 [2], the functional safety standard for the electrical, electronic and programmable electronic devices in road vehicles. Automotive SPICE and IEC/ISO 26262 thus represent the most important documents in automotive industry to comply with the Quality Management and Functional Safety assessment. The risk classification scheme defined in the IEC/ISO 26262 is called Automotive Safety Integrity Level (ASIL) and is used to rate hazards of the system based on the consideration of malfunctions in particular driving situations.

To verify the applied system, architecture, software and hardware, testing is an essential tool. It is used in order to determine whether the objectives set are met. Testing is performed based on the requirements, where the most important upgrade introduced by the IEC/ISO 26262 is the introduction of the testing of functional safety objectives. For this purpose, the ISO/IEC does not only specify the requirements regarding the processes, but also prescribes the methods which shall be applied.

Among the methods required by IEC/ISO 26262, the Hardware-in-the-Loop (HiL) testing has become indispensable in the automotive domain [8]. It represents the method where the device under test is placed in an emulated environment. The basic principle of the method is presented in the Figure 1.

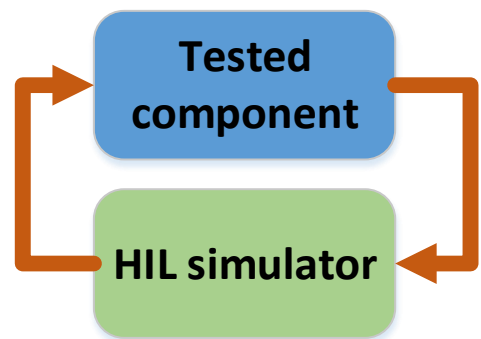

*Figure 1: HIL testing basic principle*

With the use of HiL, systems testing can be performed under close-to-real conditions. It is also important to note, that the repeatability of tests is assured, and the same testing hardware can be re-used for tests of various devices with only limited modifications. Since the device under test is available in real during testing, and no understanding of its internal design is therefore required, HiL testing is a form of black-box testing. Black-box testing can be performed for testing of units, integration testing, system testing and up to the certain level, even system validation. The testing processes as described above have been widely accepted in the automotive industry. They represent a must for any company wishing to be a part of the automotive development process or supply chain. It has been observed, however, that there is a serious lack of engineers familiar with these processes and methods, and consequently many companies, especially smaller ones, experience serious difficulties in adopting them. Definitely the knowledge in this field will be required from the future employees and therefore it is important that teaching it becomes a part of the education process, especially at higher education levels. This is especially important for the students of Electrical Engineering and Mechatronics.

The requirements for new topics and the extension of the existing ones to the education process will be presented in the following text. There have already been extensive activities in the field of adapting university curricula accordingly, in particular the Automotive Universities project [7], however those have to be extended by the testing methods, especially HiL-related ones.

This paper is organized as follows. After the introduction first the design process in the automotive industry is presented in more details. The V-model is described together with its activities. The testing in the automotive industry is described next. The main approaches, methods and processes are presented. Based on this explanation some knowledge and skills required for the HiL testing are proposed. To visualize the problem and propose a methodology of teaching, an example of testing is proposed. In the final section, the conclusion is given to summarize the most important topics required in the future education.

## II. Design Processes in Automotive industry

Automotive SPICE and the IEC/ISO 26262 provide the frameworks for development processes of embedded automotive systems. In the following we will briefly describe the essentials of both these standards.

### A. Automotive SPICE

Automotive SPICE is based on the SPICE – ISO/IEC15504 standard. It was developed for process quality management in the development of products suitable for automotive industry. In short, Automotive SPICE defines how a development process must be organized. The standard is basically a set of rules obtained from the good practices in automotive industry which were put together. Based on these practices, automotive suppliers are assessed per development project and receive the rating which determines the supplier's eligibility as a member of the supply chain for all the major global automotive OEMs, and which is usually also the basis for targeted improvement.

Figure 2 shows the V-development process that forms the backbone of both Automotive SPICE and the ISO 26262. An essential property is the bidirectional traceability between process steps on the left-hand side of the V, as well as between related process steps on the left and right-hand side of the V. While the left-hand side covers requirements engineering and design from system level to software (SW) unit level, the right-hand side is dedicated to verification and validation in the form of tests from SW unit level up to system level. In the following we will briefly describe every process step on the left-hand side of the V.

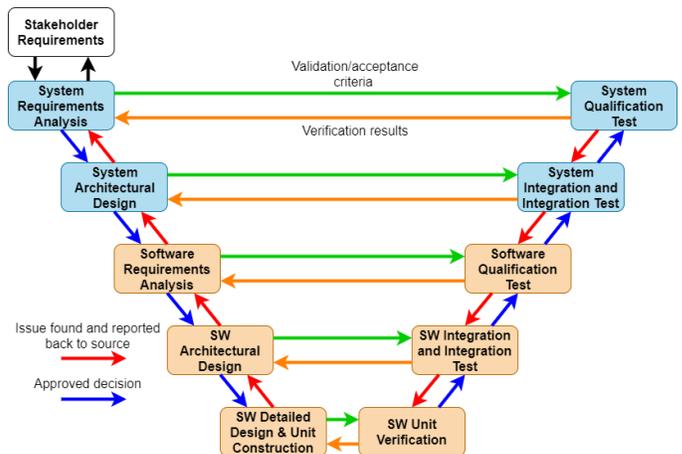
*Figure 2: V-model design (ASPICE)*

Stakeholder Requirements are (internal or external) customer requirements describing what the system shall do and how the system shall perform. The stakeholder requirements documents are analyzed with respect to the feasibility, risk and priority of requirements.

System Requirements – In the system design phase, engineers technically analyze the stakeholder requirements documents and derive company-internal system requirements from them. Each stakeholder requirement shall lead to at least one atomic (i.e., indivisible) system requirement.

The System Architectural Design maps the system requirements to system components (mechanics, electronic hardware or software) and their interfaces. At this stage, the architectural design is still a high-level black-box design specifying the main system components and their interconnections via clearly defined interfaces.

Software Requirements Analysis leads to requirements to software related parts derived from system requirements. The System Architectural design shall be used to better identify what is needed from the software.

The Software Architecture Design maps software requirements to software components and their interfaces. As in the system architectural design this is a high-level design focusing on software components and their interconnections via clearly defined interfaces.

The Software Detailed Design and Unit Construction comprises the phase for the specification and development of software units, the smallest indivisible design element. Units are the building blocks of components in the software architecture.

*B. Functional Safety (ISO/IEC 26262)*

Functional safety is becoming important in all major fields of industry. Its application together with the requirement to formalize and standardize the processes, methods and documentation resulted in creation of several standards. In the development of electric and electronic automation devices the basic standard covering functional safety is IEC/ISO 61508 [3], intended to be applicable to all fields of industry. For the specific requirements of the different industries additional functional safety standards were developed. In automotive industry, the ISO 26262 is applied (standard for road vehicles), whereas in aerospace the DO-178/ED-12 (Software Considerations in Airborne Systems and Equipment Certification) [4] and the DO-254/ED-80 (Design Assurance Guidance for Airborne Electronic Hardware) [5] are relevant. DO-178/ED-12 and DO-254/ED-80 are guidelines rather than standards, which are however treated as de-facto standards by the relevant authorities.

Functional safety does not mean that there are no faults in the final products and its operation. These will always exist, since it is a common knowledge that every system will fail at some point. The concept of functional safety means that the operation of the system will remain safe (transfer into safe state, not cause injuries or at least mitigate them) also in the presence of faults.

*C. ASPICE and Functional Safety (ISO/IEC 26262)*

ASPICE and ISO 26262 are complementary, the ASPICE process reference model can be seen in some parts of ISO 26262 chapters 4, 6, 7 and 8. Therefore, the ISO 26262 already incorporates the ASPICE process reference model (Figure 3). The ISO 26262 extends the Life Cycle Management of designing a product with safety relevant issues and adds methods to the processes.

### III. TESTING IN AUTOMOTIVE INDUSTRY

Testing is crucial in automotive industry, as ISO 26262 and ASPICE require that every requirement from stakeholder requirements to unit construction must be verified and validated. The V-model described before contains multiple testing levels and indicates the relationships between development and testing.

Problems fulfilling standardized process definitions and strictly following the V-cycle are manifold. Effort estimations for each V-cycle activity are difficult to do, especially in new developments. In most cases the number of available experienced personnel is far less than needed. Some companies use the same resources for both development and testing, which compromises the availability of specialized competences as well as the independence of developers from testers. HiL systems provide a key element for facilitating the V-cycle approach. However, using them properly requires undergoing a substantial learning curve unless engineers have prior experience from education or related professional activities.

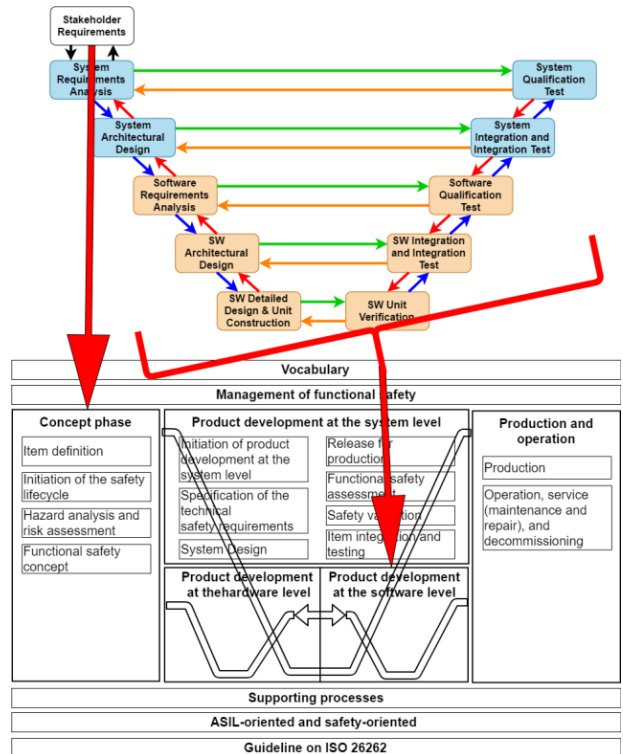

*Figure 3: ISO26262 and ASPICE implementation*

*A. Hil system*

The basic structure of a HiL system is depicted in the Figure 4. It consists of the device under test (DUT) and a HiL simulator. Both units are connected with analog and digital systems as well as busses (CAN, LIN, UART).

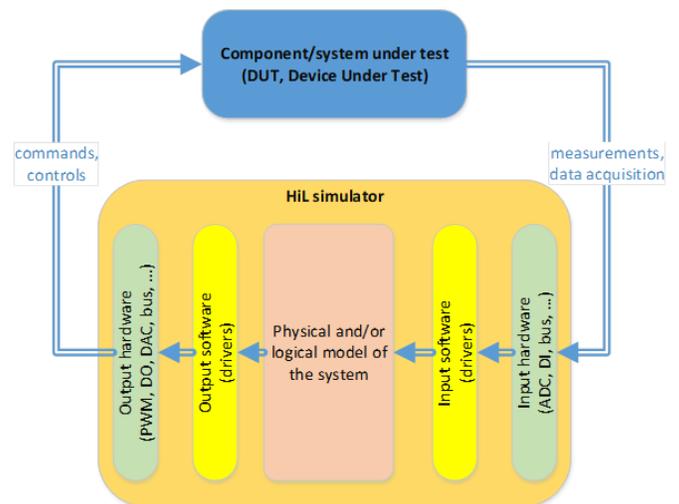

*Figure 4: HiL system*

The device under test is the system or component that needs to be tested. It can be a complete system or its part.

The HiL simulator consists of several components [7]. It is built around a system model, which represents the system's physical and/or logical behavior. Usually this is a software unit executed either on microprocessor hardware or FPGA and can be programmed with as few limitations as possible. The software for this unit can be model-based (using software like Matlab/Simulink as for HiL systems provided by dSPACE) or programmed in classical programming languages [10] [11]. The model is connected to the DUT through the input and output software (usually present in the form of drivers), which control the input and output hardware. The output hardware can consist out of PWM outputs, digital to analog converters, digital outputs and busses. The input hardware contains of analog to digital converters, digital inputs, busses, comparators, capture units etc.

The HiL simulator can be applied using the dedicated self-developed software and hardware, but because also the formal qualification is required for industrial use, commercial products from companies like dSPACE, National Instruments and Typhoon HiL are more common.

In our case, the testing was done with the help of a dSPACE HIL system that uses the model-based Simulink programming software. The device under test was a motor controller that was communicating with outside world only with bus communication. Testing is needed to verify if the DUT performs and behaves according to the stakeholder and system requirements specifications.

The test engineers shall be able to formalize every scenario of the DUT's normal operation. However, they also have to cover a wider range of scenarios than normal operation and limits of the system parameters in order to verify the system's correct and safe behavior outside the specification ranges. Such exhaustive behavior-based black-box testing is efficient, however time consuming.

Test engineer and Test Case developer shall not be the same person. Test Case developer has a wider view of the system and knows system limitations. Test engineer only knows what Test Case developer wants to test and how it should be tested. Test engineer writes code on testing equipment according to Test Case developer requirements and verifies Software/Unit requirements or validates System requirements in development. If verification/validation fails (test gives a negative result) the requirement and Test Case requirement shall be analyzed, incorrect item shall be corrected or modified to meet the desired criteria. Test Case developer and test engineer are in constant communication that involve test cases and writing tests on testing equipment - HiL system.

## IV. EXAMPLE OF DEVELOPMENT AND TESTING

Example of development and testing is represented as Application that a customer (stakeholder) is requesting a developer to develop and reproduce in the manner of ASPICE process management and part of ISO26262.

### A. Application (Stakeholder requirements document)

Basic description of functionality and what shall be done is presented to developer in as few words as possible. If there are too many demands inside one requirement, it should be broken down to more requirements. An example of stakeholder requirement (ShReqxxx, where xxx represents the requirement ID number) is given as:

*ShReq001*: *Electronic device shall measure voltage and transmit information over communication link.*

In this example it is obvious that a device shall be used that measures the voltage and transmits the information regarding its value through the communication link. The method of voltage measurement is not given, also the communication network or protocol are not defined, as this would be in a too early stage and would enforce the decision upon the developers in later stages. Namely the idea is to let them have as much freedom as possible.

### B. System Requirement

SW related requirements are derived from Stakeholders requirements according to ISO 26262 and ASPICE. In application requirement we only have one requirement, but it already indicates the need for HW and SW requirement(s) but for the purpose of presentation only one SW requirement is extracted.

*SysReq001*: *System shall monitor and transmit input voltage value over communication link with accuracy of 3% of input signal.*

Requirement is minimalistic. It shall contain system limits, but it shall not dictate final design if it is not given by the stakeholder document.

At this point it is important to define an acceptance/validation criteria. It shall contain the crucial information regarding the functionality of the system. An example can be given as:

*Measured voltage shall be communicated in the range of $U_{min}$ to $U_{max}$ with dynamic response of at least dTime.*

The values of $U_{min}$ and $U_{max}$ represent the minimal and maximal operating voltage, *dTime* represents maximum refresh time of measured value. It is important to note that symbols can be used instead of actual (numerical) values. When using symbols to give requirement flexibility, variables can be grouped in a System parameters document.

### C. Software Requirement

As first prerequisite for the frame of HW design the physical communication system or HW is chosen. Based on that decision the SW engineer can start developing SW requirements. For the purpose of presentation only one SW requirement is written here and in order to enable writing of SW requirements following HW decisions are needed. *1.) Vehicle bus standard Controller Area Network (CAN) shall be used for communication link. 2.) Translation between ADC pin and external connection shall be documented in Hardware Software Interface (HSI) document.*

*SwReq001*: *Voltage shall be represented over CAN bus communication in SI units and accuracy shall be within ±Uacc over full measured voltage operating range.*

Requirement is again minimalistic with strict limitations and demands. As it is the case in system requirements, symbols can be used, and variables can be grouped in Software parameters document.

Software requirement Acceptance/Validation criteria has the form similar as it is the case for System requirements:

*Accuracy of voltage representation over CAN bus communication shall be met. Dynamical accuracy shall be considered according CAN to bus communication delay.*

### D. Software Architecture

Software architecture is defined on a higher level than Software Detailed Design and Unit Construction (later stage) and this shall be respected. It shall roughly define Software Requirement functionalities, interfaces and timing constrains. Software architecture shall group Software functionality that covers all SW requirements.

SW Architecture is already described in Software requirements, but there it is not represented yet in the graphical form. Architecture is developed in a graphic way and represents a starting point of Software Detailed Design and Unit Construction stage.

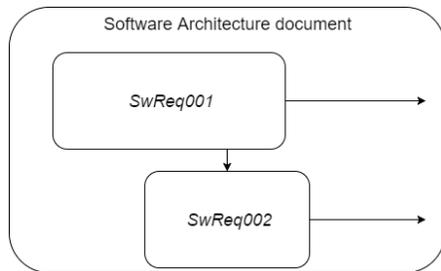

*Figure 5: Simplified Software Architecture document*

Simplified Software Architecture document (Figure 5) is used as a basis for SW integration strategy, but it can also be used for overview of the SW Qualification Test plan.

### E. Software Detailed Design and Unit Construction

Software Detailed Design shall be chosen by analyzing Software requirements and Software Architectural design document. Great care shall be taken in data flow and what functions are time critical. Several Software Units can be derived from single Software requirement.

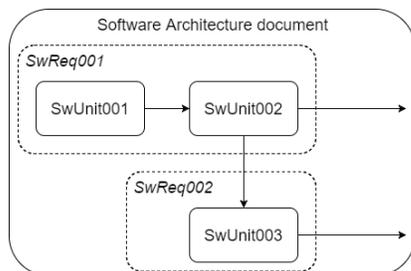

*Figure 6: Simplified Software Architecture with SW Units development*

Dashed line frame in (Figure 6) indicates which Software Units belongs to which SW requirement. It shows that at least one Software Unit can be extracted from one Software requirement.

*SwUnit001: Voltage on the pin named VDC shall be acquired using the driver for the voltage measurement with ADC every 100 μs. Output value named U_DC shall represent voltage value in SI units.*

*SwUnit002: U_DC value shall be sent over CAN bus communication according CAN communication document.*

### F. Test Case

Tests are required in order to verify or validate the correct operation or design of the device. For that purpose, the Test Cases (TC) need to be developed. Format of Test Case shall be in a form that can be easily reproduced, repeated or copied. Test Case should be easy and simple to understand.

An example of test case is given below:

*TC001:*
<u>*Verify*</u> *voltage accuracy represented over CAN bus communication.*
<u>*Using*</u> *HiL system to communicate and acquire voltage information over CAN bus communication.*
<u>*With*</u> *changing DUT DC-link voltage over full-scale operating range.*
<u>*To*</u> *verify if supply voltage on DUT DC-link and represented voltage over CAN bus communication meet voltage accuracy of at least ±0.5V. Dynamical accuracy shall be considered according to CAN bus communication delay.*

Verify, Using, With, To form[13] is used in above example for description section to make it easy to understand. As process develop the consistent usage of this form makes it easier for test engineer to understand what test case developer was thinking. "Verify" is used as a first word in test statement, explaining what will be tested/verified. "Using" defines the applied tools and devices. Sentence that starts with "With" gives the procedure/method of test and sentence that starts with "To" gives information about the expected results of the success of the verification.

### G. Test (steps)

To perform the test as required by Test Case, the following steps must be performed in the presented sequence:

*1. Initialize DUT – power up DUT with normal operating voltage,*

*2. Pool power supply voltage and voltage represented over CAN bus communication,*

*3. Change power supply voltage to minimal operating voltage value and start incrementing voltage in steps of 0.1V every 1s and finish at maximum operating value,*

*4. Voltage from DC-link power supply and voltage represented over CAN bus communication shall not deviate for more than ±0.5V,*

*5. No error massages shall be triggered during full operating voltage change*

This text also must be a part of the Test Case document.

## V. Knowledge and Skills Required for HiL Testing

More and more educational institutions teach how teamwork is important in everyday life but practice independent work. In fact, teamwork is something that research & development (R&D) industries are seeking to achieve. It is known that most experts are prone to bad communication and poor choice of words at crucial times (public appearances). Learning to communicate in the team and exchange the data clearly is essential. Ability to function in the team is highly appreciated for an engineer.

Basic knowledge about the operation of the system is required. Likewise, the use of semi-formal modelling methods, like UML (Universal Modelling Language) or similar, is very valuable. In the future it is expected that formal methods will be required.

System engineering is an important skill. The device to be designed needs to be understood as a system and details should be avoided in early stages, because they can lead to limiting the design possibilities. The design needs to remain open to different solutions and at the earlier stages the details don't have to be solved. The urge for and engineer to solve the problems too early in many cases leads to the problems in performing the processes required by ISO 26262 and ASPICE. Actually, those two guidelines/standards incorporate hierarchy as the main methodology in the design of the system. The roles of the involved personnel must be clearly divided and clarified. This gives all the participants their tasks and enables them to act individually, interacting based on clear instructions. Thus, each participant can check the predecessor in the process and give a clear output to the developers of the following stages, not limiting them in their work[9].

Knowledge about hardware design and its limitations is required for the development of hardware requirements and testing. Software developers, more precisely the software engineers developing hardware related software (drivers, etc.), also require knowledge about the processor and its peripherals. Especially it is important to note that a hardware software interface (HSI) is required and represents an essential part of design.

### A. Application

Application information is typically provided by customer especially if the device to be designed is a component of the system. In other case (if complete system is designed) the general knowledge about system operation is required.

### B. System Requirements

For the purpose of system requirements definition, the knowledge regarding system operation needs to be more technical. Knowledge from the fields of system engineering and quality management need to be added to the general technical knowledge but detailed knowledge regarding applied hardware and software is not required.

### C. System Architecture

To map the system requirements to the system components knowledge from the fields of system, hardware, software and mechanical engineering need to be used to develop a good system platform that is partitioned into many sub components that can be developed separately. Basic engineering skills are required, but the main issue is to be able to understand interaction among the subcomponents. This could be a typical position for a mechatronics engineer.

### D. Software Requirements

When writing software requirements expert knowledge is needed about software tools to be applied. Basic knowledge about hardware items is needed in order to understand how the hardware is constructed and interfaced with outside world. Functionality of hardware shall be well known to the software engineer that writes Software Requirements.

### E. Software Architecture

A higher level of approach is needed in Software Architecture development. To form architecture from Software requirements, knowledge of system and software engineering need to be known. This would be typically a task performed by a software engineer.

### F. Software Detailed Design and Unit Construction

To develop a detailed design from Software architecture and Software requirements expert knowledge is needed about software tools. To develop drivers higher understanding of the hardware setup is needed. In case of the more hardware-oriented units this task could be performed by an embedded software engineer, for more general units a knowledge of programming should be sufficient.

### G. Test Case development

To write test cases knowledge about quality management and consistency is a key for success. Before starting to write the Test Case, System, Software, Hardware and Architecture related documents for whole System under test are needed. The possibilities of the testing equipment and software must be well known to the test case developer. Hardware and Software Interfaces shall be well defined. Schematic diagram of DUTs Hardware shall be available to tester and all DUT interfaces to the outside world need to be described in high detail. In order to be able to perform these tasks, test case developer should understand the operation of the system or unit for which the test case shall be written. The internal structure and software should not be known to the test case developer, because the test cases need to be developed from requirements and not from the design. Therefore, the test case developer should possess the knowledge from the field of system engineering and be able to utilize the available testing tools to recreate the input/output behavior. The test engineer should be able to use the testing tools, which requires some programming skills (in the programming tool used in testing devices) and ability to connect the DUT to the testing device. The role of test case developer should be given to an experienced engineer whereas the role of test engineer can be assumed by less experienced personnel, giving them a good possibility to learn about the processes and devices.

### H. HiL Test programing

Expert knowledge is needed about software tools used by the testing equipment especially if advanced HiL system is used for testing software and system requirements. Expert knowledge is needed about limitations of testing hardware equipment. Test programmer must be able to predict what test cases can be executed and what test cannot be tested with available equipment. Model-based approaches and tools are essential for the effective use of the modern HiL equipment.

To be able to develop more complex HiL tests the knowledge about the system modelling and simulation is required. The HiL test developer needs to be able to develop the simulated environment for the DUT, which typically consists of the physical and logical model of the DUT environment, together with the interfaces. For that purpose, at least basic knowledge in the field of electrical engineering is required. Control engineering is also essential, because the models in HiL need to be controlled in order to emulate the behavior of an actual environment. Modelling and identification are highly useful tools.

## CONCLUSION

Based on the findings described in the previous text and experiences gathered in the projects dealing with the development of the safety critical applications in the automotive industry it can be determined that besides the required knowledge in the specific field of study, which the student should acquire in the frame of education process, should also contain the knowledge of software engineering, especially methods and processes of design and testing. More specific, this knowledge includes some essential contents from the requirements engineering, design of experiments, software engineering and system engineering. The typical skills required include the use of Application Lifecycle Management (ALM) tools (like Polarion, Doors, etc.) and use of version-control systems (Git, SVN). It is not very important which tools are used, the main goal is to make students familiar with the fact that they exist and understand their basic meaning and function. Model-based approaches (like Universal Modelling Language, UML) are also important, as they are gaining popularity in the automotive industry, significantly reducing the time-to-product and reducing the possibilities of mistakes in the development process, not only due to faults in programming, but also in failures to understand the specifications and requirements. Tools like Matlab/Simulink are becoming classical choice in the development process. In future even advanced approaches like formal methods will become important.